\documentclass{elsart}

\usepackage{epsfig}
\usepackage{graphicx,amsfonts,amsmath,eucal}
\usepackage{amssymb}

\begin{document}

\begin{frontmatter}

\title{Multireference
configuration interaction treatment of excited-state electron correlation
in periodic systems: the band structure of
{\bfseries \itshape trans}-polyacetylene}

\author{Viktor Bezugly and}
\author{Uwe Birkenheuer\corauthref{cor1}}
\ead{birken@mpipks-dresden.mpg.de}

\corauth[cor1]{Corresponding author. Fax: ++49 351 871-2199.}

\address{Max-Planck-Institut f\"ur Physik komplexer Systeme,
N\"othnitzer Str. 38, 01187 Dresden, Germany}

\begin{abstract}
A systematic method to account for electron correlation in periodic
systems which can predict quantitatively
correct band structures of non-conducting solids from first
principles is presented.
Using localized Hartree-Fock orbitals
(both occupied and virtual ones), an effective Hamiltonian is built up whose
matrix elements can easily be transferred from finite
to infinite systems.
To describe the correlation effects wave-function-based
multireference configuration
interaction (MRCI) calculations with singly and doubly excited
configurations are performed. This way it is possible to generate,
both, valence and conduction bands with all correlation
effects taken into account.
{\it trans}-polyacetylene is chosen as a test system.
\end{abstract}

\end{frontmatter}

\section{Introduction}
\label{1.}

Recently, with the rapid progress in computing facilities,
rather sophisticated numerical methods started to arise
which attempt to determine quantitatively correct electronic
band structures of solids from first principles. These
methods can be grouped in two families: those focusing on 
ground-state properties such as the electron density (or the 
quasi-particle Green function) and those trying to approach
the full many-body electron wave function.

As a rule, traditional density functional methods are used
in solid state physics to study the electronic properties
of infinite periodic systems, but since these approaches are
not designed for excited states they usually strongly underestimate
the band gaps of non-conducting systems. Also these methods
do in general not allow improvements in some controlled way.

On the other hand, methods based on many-body wave functions
of finite systems are well-established in quantum chemistry. They
usually start from an approximate wave function
(e.g.\ the Hartree-Fock (HF) wave function) and approach the 
correct solution by improving the wave function systematically.
The development of such methods for solids only started recently
(see below).
The band gaps obtained on the HF level (via Koopmans' theorem)
are usually far too large, but the situation improves 
steadily upon successive inclusion of correlation effects.
In this sense the wave-function-based methods are more general
though more demanding in predicting quantitatively
correct electronic structures of periodic system than
the common density functional methods.

Among the recently developed wave-function-based approaches
to correlated band structures one can distinguish those
which employ local Hamiltonian matrix elements
in real space \cite{Graef93,Graef97,Albrecht00,Albrecht02},
those which work directly in $k$-space and use
many-body perturbation theory to account for electron
correlation \cite{Sun96,Ayala01}, and those which
use an approximate Green function formalism
\cite{Albrecht02,Albrecht01,Albrecht02b}.
Other approaches to electron correlation in solids
\cite{Pisani04,Buth04} are presently implemented,
corroborating the currently high interest 
in this new and rapidly growing research field.

The aim of this Letter is to present the local Hamiltonian method
which originates from the approach to valence bands described in
Ref.~\cite{Graef93,Graef97,Albrecht00} but which can now also
treat conduction bands and thus allows to predict {\it complete}
band structures (and in particular band gaps) of non-conducting
systems with controlled accuracy.
The presented method is designed to be used with any suitable 
standard quantum-chemical program package. We decided for the 
multireference
version of the single and double configuration interaction
method (MRCI(SD)) to account for electron correlation in
both the ground and excited states of the system
together with a special size-consistency correction we have developed
for that purpose \cite{disser}.
This explicitly includes up to five-particle configurations (three holes
and two electrons or vice versa) and therefore goes far beyond any of
the other wave-function-based methods mentioned above.

To achieve our goal, the original problem of determining delocalized
electronic states with a crystal momentum (Bloch states) is reformulated
in terms of localized orbitals in real space (Wannier functions).
In this representation one can take advantage of the predominantly 
local character of correlation effects and evaluate them step-by-step
in finite clusters of the periodic system.
The approach is purely {\it ab initio}. We adopt
the experimentally known geometry of our test system
({\it trans}-polyacetylene) and we do not make any approximation
for the electron-electron or electron-nuclear interaction.

\section{The method}
\label{2.}
\newcommand{\kkvec}{{\mbox{\normalsize \boldmath $k$}}}
\newcommand{\RRvec}{{\mbox{\normalsize \boldmath $R$}}}
\newcommand{\kvec }{{\mbox{\scriptsize \boldmath $k$}}}
\newcommand{\Rvec }{{\mbox{\scriptsize \boldmath $R$}}}
\newcommand{\Zvec }{{\mbox{\scriptsize \boldmath $0$}}}

The valence band energies $\varepsilon_{\kvec\nu}$ of a periodic
system are defined as
\begin{equation}\label{val_band}
  - \varepsilon_{\kvec\nu} = {\rm IP}_{\kvec\nu}
                           = E^{N-1}_{\kvec\nu} - E^N_0
\end{equation}
where $E^N_0$ is the ground-state energy of the neutral $N$-electron system
and $E^{N-1}_{\kvec\nu} = \langle \Psi^{N-1}_{\kvec\nu\sigma} | H |
                          \Psi^{N-1}_{\kvec\nu\sigma} \rangle$
is the energy of that excited state $\Psi^{N-1}_{\kvec\nu\sigma}$
of the ($N$-1)-electron system
which corresponds to a quasi-hole of spin $-\sigma$, crystal momentum
$-\kkvec$ and valence band index $\nu$.
The corresponding equation for the conduction band energies
$\varepsilon_{\kvec\mu}$ reads
\begin{equation}\label{cond_band}
  - \varepsilon_{\kvec\mu} = {\rm EA}_{\kvec\mu}
                           = E^N_0 - E^{N+1}_{\kvec\mu}
\end{equation}
with $\mu$ labeling the conduction bands
and $E^{N+1}_{\kvec\mu} = \langle \Psi^{N+1}_{\kvec\mu\sigma} | H |
                          \Psi^{N+1}_{\kvec\mu\sigma} \rangle$
referring to the excited-state of the ($N$+1)-electron
system which is associated with a quasi-particle of spin $\sigma$ in a 
virtual Bloch orbital $\varphi_{\kvec\mu}$.
Note, that the band energies are differences of {\it total} energies rather
than simple one-particle energies. Also note, that excited
states which do not have any counter part in the single-orbital
picture such as satellite and shake-up states are explicitly 
excluded here.

On the Hartree-Fock level the ground-state wave function of the neutral
system is approximated by a single Slater determinant $\Phi$, and 
because in infinite systems there is no orbital relaxation upon removal
or adjunct of a single electron, Koopmans' theorem holds such that the
excited states of the ($N$$\pm$1)-electron system are simply given by
\begin{equation}\label{PhiN+-1k}
  | \Phi^{N-1}_{\kvec\nu\sigma} \rangle
  = c_{\kvec\nu\sigma} | \Phi \rangle
  \quad\mbox{and}\quad
  | \Phi^{N+1}_{\kvec\mu\sigma} \rangle
  = c^{\dagger}_{\kvec\mu\sigma} | \Phi \rangle
  \quad.
\end{equation}
Here $c_{\kvec\nu\sigma}$ and $c^{\dagger}_{\kvec\mu\sigma}$ create or
annihilate an electron in the respective HF spin orbitals
which can directly be obtained (together with the HF band energies)
from standard HF program packages for periodic systems like
CRYSTAL \cite{CRYSTAL}.

\begin{figure}
\centerline{
\psfig{figure=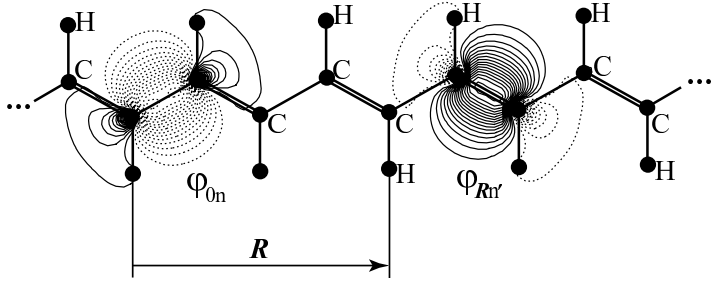,height=3.7cm}
}
\caption{Two typical local orbitals in {\it trans}-polyacetylene. Contours are
         in steps of 0.025 au.}
\label{tPA}
\end{figure}

Applying separate multi-band Wannier transformations to the occupied and
virtual Bloch orbitals provides two sets of local Wannier functions,
$\varphi_{\Rvec n}$ and $\varphi_{\Rvec m}$, where $\RRvec$ denotes the
lattice vector to the unit cell associated with each Wannier function (see
Fig.~\ref{tPA}).
Using these Wannier functions so-called local
one-particle configurations
\begin{equation}\label{PhiN+-1R}
  | \Phi^{N-1}_{\Rvec n\sigma} \rangle
  = c_{\Rvec n\sigma} | \Phi \rangle
  \quad\mbox{and}\quad
  | \Phi^{N+1}_{\Rvec m\sigma} \rangle
  = c^{\dagger}_{\Rvec m\sigma} | \Phi \rangle
\end{equation}
can be introduced, where $c_{\Rvec n\sigma}$ and $c^{\dagger}_{\Rvec m\sigma}$
create or annihilate an electron from the respective Wannier spin orbital.
Having switched to the local configurations the Hartree-Fock band energies can
be recovered by diagonalizing the $k$-dependent 'ionization potential' and
'electron affinity' matrices,
\begin{eqnarray}\label{IPn}
  {\rm IP}^{\rm HF}_{nn^\prime}(\kkvec)
  &=&\sum_\Rvec {\rm e}^{i\kvec\Rvec} \: {\rm IP}^{\rm HF}_{nn^\prime}(\RRvec)
  \quad\mbox{with}\nonumber\\
  {\rm IP}^{\rm HF}_{nn^\prime}(\RRvec)
  &=&\langle \Phi^{N-1}_{\Zvec n\sigma} | H |
             \Phi^{N-1}_{\Rvec n^\prime\sigma} \rangle
   - E^N_0 \, \delta_{\Zvec R} \delta_{nn^\prime}
\end{eqnarray}
and
\begin{eqnarray}\label{EAm}
  {\rm EA}^{\rm HF}_{mm^\prime}(\kkvec)
  &=&\sum_\Rvec {\rm e}^{i\kvec\Rvec} \: {\rm EA}^{\rm HF}_{mm^\prime}(\RRvec)
  \quad\mbox{with}\nonumber\\
  {\rm EA}^{\rm HF}_{mm^\prime}(\RRvec)
  &=&E^N_0 \, \delta_{\Zvec R} \delta_{mm^\prime}
   - \langle \Phi^{N+1}_{\Zvec m\sigma} | H |
             \Phi^{N+1}_{\Rvec m^\prime\sigma} \rangle
  \quad.
\end{eqnarray}

The same holds for the {\it correlated} band energies provided the Wannier
transformations which mediates between the Bloch and the Wannier
orbitals is applied to the excited-state wave functions as well, yielding 
\begin{eqnarray}
  \label{Psi-rep}
  \Psi^{N-1}_{\kvec\nu\sigma}&=&\sum_n U_{\nu n}(\kkvec)
                                \sum_\Rvec {\rm e}^{-i{\kvec\Rvec}} \:
                                \Psi^{N-1}_{\Rvec n\sigma}
  \quad\mbox{and}\\
  \label{Psi+rep}
  \Psi^{N+1}_{\kvec\mu\sigma}&=&\sum_m U^\ast_{\mu m}(\kkvec)
                                \sum_\Rvec {\rm e}^{i{\kvec\Rvec}} \:
                                \Psi^{N+1}_{\Rvec m\sigma}
  \quad,
\end{eqnarray}
where $U_{\nu n}(\kkvec)$ and $U_{\mu m}(\kkvec)$ are the band-mixing
matrices of the two Wannier transformations.
The {\it correlated} ionization potential and electron affinity matrices
${\rm IP}_{nn^\prime}$ and ${\rm EA}_{mm^\prime}$ are
defined in perfect analogy to Eqs.~(\ref{IPn}) and (\ref{EAm}).

The transformed many-body wave functions $\Psi^{N-1}_{\Rvec n\sigma}$ and 
$\Psi^{N+1}_{\Rvec m\sigma}$ are sometimes referred to as 
'correlated local holes' and 'correlated local electrons'.
Yet, like the Wannier functions or any other kind of localized orbitals they
are pure mathematical constructions only meant to provide a convenient
{\it local} representation of the delocalized eigenstates
$\Psi^{N-1}_{\kvec\nu\sigma}$ and $\Psi^{N+1}_{\kvec\mu\sigma}$
one is looking for.

The benefit of this local representation is two-fold. Firstly, the
off-diagonal (or 'hopping') matrix elements of ${\rm IP}_{nn^\prime}(\RRvec)$
and ${\rm EA}_{mm^\prime}(\RRvec)$ decay rapidly with increasing distance
$\RRvec$ when well-localized Wannier orbitals are used
such that the (in fact infinite) summation over $\RRvec$ in Eqs.~(\ref{IPn})
and (\ref{EAm}) can be truncated at some distance $R_{\rm cut}$.
Secondly, the individual local matrix elements ${\rm IP}_{nn^\prime}(\RRvec)$
and ${\rm EA}_{mm^\prime}(\RRvec)$ can be evaluated in {\it finite}
clusters of the periodic system, because excitations from the 
involved local spin orbitals into some distant spin orbitals
become negligible once they are sufficiently apart. In addition, 
as outlines in Ref.~\cite{Graef97,Albrecht00}, asymptotic corrections can
be applied to account for the missing contributions, if necessary.

To obtain the local matrix elements (LMEs) from the clusters our local
Hamiltonian approach is applied to the finite cluster as well leading to a set 
of localized occupied and virtual orbitals $\varphi_a$ and $\varphi_r$
and the corresponding molecular ionization potential and electron
affinity matrices ${\rm IP}_{aa^\prime}$ and ${\rm EA}_{rr^\prime}$.
However, instead of determining
the LMEs in order to find the band energies as done
for the periodic system, one now proceeds in the opposite direction. Starting
from the excited-state energies $E^+_\nu$ and $E^-_\mu$ of the cationic and
anionic cluster, which can be calculated (together with the molecular 
ground-state energy $E_0$) by means of any suitable quantum chemical
correlation method, the unknown LMEs are evaluated according to
\begin{eqnarray}
  \label{IPa}
  {\rm IP}_{aa^\prime}\,&=& \sum_\nu E^+_\nu \: U_{\nu a} U^\ast_{\nu a^\prime}
                         -  E_0\,\delta_{aa^\prime}
  \quad\mbox{and}\\
  \label{EAr}
  {\rm EA}_{rr^\prime}  &=& E_0\,\delta_{rr^\prime}
                         -  \sum_\mu E^-_\mu \: U^\ast_{\mu r} U_{\mu r^\prime}
  \quad.
\end{eqnarray}
The unitary matrices $U_{\nu a}$ and $U_{\mu r}$ stem from some suitable 
localization such as the Foster-Boys procedure \cite{Boys60}
(see Fig.~\ref{tPA}) and are the molecular counter parts of the band-mixing
matrices.
We used the MRCI(SD) option \cite{Werner88,Knowles88,Knowles92} of the MOLPRO
program package \cite{MOLPRO}
taking the unrelaxed Hartree-Fock orbitals of the neutral system to expand 
the wave functions of the charged molecules as well.
The lack of size-consistency (SC) inherent to truncated CI schemes
could be overcome by developing a special SC correction
for the ($N$$\pm$1)-electron systems \cite{disser} similar to
the well-established Pople correction
\cite{Pople77} for closed-shell systems.
To facilitate the still quite demanding cluster calculations the incremental
scheme was employed (see Ref.~\cite{Graef97,Albrecht00,disser} for details).

Despite the formal hole-particle dualism in the equations above, there
is quite a difference between valence and conduction band 
calculations for {\it ab initio} Hamiltonians.
Virtual orbitals are much harder to localize
than occupied ones and the anionic clusters tend to become electronically
unstable such that some embedding or confinement techniques have to be applied
to make the calculations feasible.

\section{Results}
\label{3.}

As test system for our method we chose a
{\it trans}-polyacetylene single chain, an infinite, flat
hydro-carbon molecule with alternating C--C bond lengths
and C$_2$H$_2$ as repeat unit (see Fig.~\ref{tPA}).
The experimental geometry of the carbon skeleton \cite{Kahlert87},
$d$(C--C) = 1.45{\AA} and $d$(C=C) = 1.36{\AA} with a lattice constant 
of 2.455{\AA}, is adopted resulting in a bond angle
$\angle$(C--C,C=C) of 121.74$^\circ$, and the C--H bond length is 1.087{\AA}
(taken from Ref.~\cite{Suhai92}).
Flexible correlation-consistent valence triple zeta (VTZ) basis sets
\cite{Dunning89} were used throughout, to allow
reasonable correlation calculations also for the anionic
($N$+1)-electron states.

X-ray scattering tells \cite{Fincher82} that crystalline
{\it trans}-polyacetylene (tPA) consists of weakly coupled fishbone-like
packed tPA single chains all aligned in the same direction, as is also
evident from the high anisotropy of the electronic properties of bulk tPA.
Our HF calculations with CRYSTAL give valence band widths of 5-10 eV
for crystal momenta $\kkvec$ parallel to the chains but not more 
than 0.4 eV in the perpendicular directions. The same holds for the low-lying
conduction bands up to about 3 eV above the lowest conduction band energy.
Hence, up to this energy, bulk tPA can, in fact, be considered as a quasi
one-dimensional crystal and
a single tPA chain with the finite basis sets is
an appropriate model for investigating that part of the electronic structure.

\begin{figure}
\centerline{
\psfig{figure=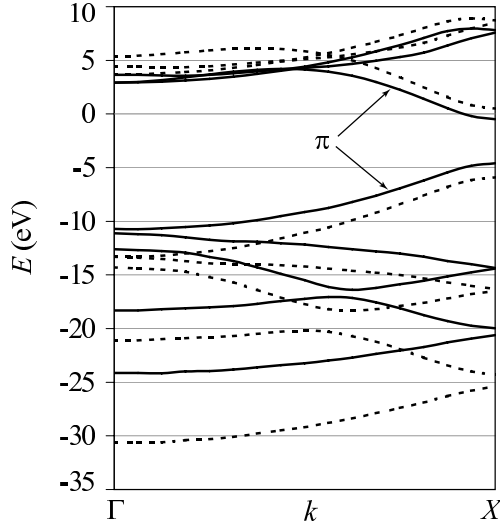,height=7cm}
}
\caption{Correlated (solid line) and Hartree-Fock (dashed line) band structure
of an infinite {\it trans}-polyacetylene single chain.}
\label{bands_fig}
\end{figure}

The band structure of a single tPA chain as obtained by our local Hamiltonian
approach is depicted in Fig.~\ref{bands_fig} where it is compared
to the HF data. All five valence bands and the three lowest conduction
bands of tPA are shown. Evidently, the band structure changes dramatically
upon inclusion of electron correlation. The valence bands shift
upwards, the conduction bands downwards, and a flattening of all bands
(significantly more pronounced in the valence region) is observed leading to
the desired narrowing of the band gap.
Other interesting but more
subtle changes are also discernible, e.g. the lifting of accidental degeneracies
or the narrowing of avoided crossings.
The band gap of tPA reduces by 2.31 eV from 6.42 eV on the HF level down
to 4.11 eV when correlation is included.
This value is perfect in line with
similar values, 2.35 and 2.38 eV, reported in two independent studies
on tPA single chains
\cite{Sun96} and \cite{Ayala01}, where the $\pi$ bands alone were
investigated by means of M{\o}ller-Plesset perturbation theory (MP2).
The lowest ionization potential becomes 4.58 eV, the highest electron affinity
0.47 eV, a positive value finally, indicating
that {\it trans}-polyacetylene is able to retain an extra electron
as is known from experiment \cite{Kaner89}.

%
%

The accuracy of our approach is governed by the error in the
determination of the LMEs by the incremental scheme. For tPA it mainly arises
from the truncation of the incremental series after the second order
corrections. The discarded long range polarization effects
turned out to be negligible.
The resulting uncertainty in the band structure could
be estimated to not exceed 0.3 eV for any $k$-point
which is less than 1\% of the total band energy range.
A more detailed error analysis can be found elsewhere \cite{disser}.
The new size-consistency (SC) correction for the excited states
is crucial here to reduce the SC error in the LME increments to 0.01\% of the 
contributing correlation energies.
The truncation error of the lattice summations in Eqs.~(\ref{IPn})
and (\ref{EAm}) was analyzed on the HF level. Keeping only
LMEs above 1 mHartree (which results in 59 LMEs for the valence and 39
for the conduction bands) the band energies from a canonical HF calculation
with CRYSTAL could be reproduced within 0.2 eV.
These deviations can easily be accounted for by
adding them to the band energies emerging from the LMEs
(as done in Fig.~\ref{bands_fig}).

\section{Discussion and Conclusions}
\label{4.}

In optical studies on bulk {\it trans}-polyacetylene
\cite{Fincher82,Tani80,Leising88} an absorption maximum
at about 2 eV is observed. Taking into account the
evidence, both experimentally and theoretically
for the existence of an excitonic gap state
\cite{Sun96,Friend99,Rohlfing99,Molinari02},
0.4 eV below the conduction band \cite{Rohlfing99},
the fundamental band gap of tPA becomes 2.4 eV,
which is noticeably smaller than the value of
4.1 eV we (and others \cite{Sun96,Ayala01}) found.
$\pi$-band splitting due to interchain coupling was made
responsible for that discrepancy \cite{Ayala01}, but for
the experimentally found fishbone-like stacking of bulk tPA
we only find a reduction of the fundamental band gap of
0.6 eV on the HF level and suppose (based on the MP2 data in
Ref.~\cite{Ayala01}) a further 0.2 eV reduction upon 
inclusion of correlation leading to a gap of 3.3 eV.
The more pronounced reductions ($\sim 1.4$ eV) reported in
Ref.~\cite{Ayala01} for two tPA chains are considered to be an artifact of the
vertical tPA packing assumed there.

The missing long-range {\it interchain} polarization existing in the bulk material
could be the reason why wave-function-based methods
applied to tPA single chains (or stacks of a few of them) consistently
overestimate the fundamental band gap of tPA, but it 
could also be that effects beyond bare electron correlation
have to be included to achieve accordance with
experiment.

Nevertheless, this study demonstrates that our local Hamiltonian
approach provides a sought scheme to perform reliable 
wave-function-based correlation calculations for excited
electron hole and attached-electron states in solids and
polymers with any suitable quantum chemical correlation
method available, as corroborated by similar studies in our
group on diamond and hydrogen fluoride chains.

\section{Acknowledgements}

We thank Professor P.~Fulde for the fruitful discussions.
This work was partially supported by the Deutsche Forschungsgemeinschaft
via SPP 1145.

\end{document}